\documentclass[aps,prx,reprint,superscriptaddress, twocolumns]{revtex4-2}
\usepackage{amsmath,amssymb}
\usepackage{bbold}
\usepackage{float}
\usepackage{graphicx}
\usepackage{dcolumn}
\usepackage{bm}
\usepackage[colorlinks]{hyperref}
\usepackage{braket}
\usepackage{mathtools}
\usepackage{comment}
\usepackage{xcolor}
\usepackage{physics}
\usepackage{enumerate}

\colorlet{alena}{blue!80!gray}
\colorlet{roman}{green!40!gray}

\begin{document}

\title{Comparison between Tensor Networks and Variational Quantum Classifier}

\author{G.~Laskaris}
\affiliation{Terra Quantum AG, Kornhausstrasse 25, 9000 St. Gallen, Switzerland}
\affiliation{LIACS, Leiden University Leiden, Netherlands}

\author{Ar.~A.~Melnikov}%
\affiliation{Terra Quantum AG, Kornhausstrasse 25, 9000 St. Gallen, Switzerland}

\author{M.~R.~Perelshtein}%
\affiliation{Terra Quantum AG, Kornhausstrasse 25, 9000 St. Gallen, Switzerland}

\author{R.~Brasher}%
\affiliation{Terra Quantum AG, Kornhausstrasse 25, 9000 St. Gallen, Switzerland}

\author{F.~Neukart}
\affiliation{Terra Quantum AG, Kornhausstrasse 25, 9000 St. Gallen, Switzerland}
\affiliation{LIACS, Leiden University Leiden, Netherlands}

\author{T.~Bäck}%
\affiliation{Terra Quantum AG, Kornhausstrasse 25, 9000 St. Gallen, Switzerland}
\affiliation{LIACS, Leiden University Leiden, Netherlands}

\begin{abstract}
The primary objective of this paper is to conduct a comparative analysis between two Machine Learning approaches: Tensor Networks (TN) and Variational Quantum Classifiers (VQC). While both approaches share similarities in their representation of the Hilbert space using a logarithmic number of parameters, they diverge in the manifolds they cover. Thus, the aim is to evaluate and compare the expressibility and trainability of these approaches. By conducting this comparison, we can gain insights into potential areas where quantum advantage may be found. 
Our findings indicate that VQC exhibits advantages in terms of speed and accuracy when dealing with data, characterized by a small number of features. However, for high-dimensional data, TN surpasses VQC in overall classification accuracy. We believe that this disparity is primarily attributed to challenges encountered during the training of quantum circuits. 
We want to stress that in this article, we focus on only one particular task and do not conduct thorough averaging of the results. 
Consequently, we recommend considering the results of this article as a unique case without excessive generalization.
\end{abstract}

\maketitle

\section{\label{sec:introduction} Introduction}

A quantum computer encodes information in quantum states and utilizes different quantum mechanical properties, such as superposition, entanglement, and interference to execute calculations \cite{QuanCOmpDef, Chuang, QC0, QC1, Marinescu, Shor1, Shor2}. Quantum machine learning (QML) is machine learning implemented in a quantum computer or a quantum simulator. QML can be realized, for example, by parameterized quantum circuits (PQC). A specific flavor of PQC that we discuss in this paper is variational quantum classifiers (VQC). VQC models are PQC which are optimized using classical optimization algorithms to find the optimal value of the parameters \cite{cerezo_variational}.  

Related to QML techniques are quantum-inspired algorithms such as tensor network (TN) models \cite{quantum_state_preparation_TQ, TN_explanation_paper_2,TTN-VQC_similar_paper}. Typically, the way that VQC and TN operate, are similar. Their main difference is the way they manipulate the weights of the model (see Sec.~\ref{sec:Implementation}). TN utilizes the TT-decomposition to manipulate efficiently the high-dimensional tensors that represent the weights of the model and the features of the data. The parameters that consist the weights tensor of TN are optimized using the Riemannian optimization algorithm \cite{riemannian,Riemannian_Manifolds,Exponential_Machines}.

In this work, we apply VQC and TN to publicly available datasets to compare the performance of models. We are following up on research where the performance of the TN model was compared to that of standard gradient descent (SGD) in classification tasks \cite{Exponential_Machines}. There, the superiority of the TN model over SGD was shown for different datasets.

Among the many applications of the models used in this paper are applications of QML and quantum-inspired tensor networks in high-energy physics \cite{QML_and_high-energy_physics,QML_supremacy_high-energy_physics,QML_Cern, TN_high-energy_physics}. Moreover, applications in image classification \cite{image_classification,Mnist_vqcddn,VQCTN_iris_mnist} can be used in autonomous systems of self-driving cars and unmanned aerial vehicles \cite{automotive}. The need for powerful computations is evident in these applications since they require decision-making in real-time, along with fast adaptation to the environment. Those applications strongly indicate that QML produces impressive results in comparison to classical models.

In general, QML is a promising candidate for solving demanding tasks in various fields. For example, in drug discovery, it is proven that QML techniques, such as generative adversarial networks (GAN) and convolutional neural networks (CNN) are superior to their classical analog \cite{drug_discovery, QC1, drug2}. Other applications of QML models lies within the spectrum of Reinforcement Learning (RL) \cite{satelite_application, RL}.
The main goal of this work is to give a more solid understanding of different ML models \cite{Applications_of_ML}. Specifically, we investigate the respective strengths and weaknesses of a variational quantum classifier and a tensor network model.

This article is structured as follows. In Sec.~\ref{sec:preliminaries}, we provide a short introduction to all necessary preliminaries to grasp the essence of this research. The preliminaries include a brief introduction to machine learning, in Sec.~\ref{sec:ML}, and tensor networks, in  Sec.~\ref{sec:TN}. There, important techniques will be introduced, namely principal component analysis (PCA) \cite{PCA, dimensionality_reduction_survey}, Variational circuits \cite{VQC,quantum_state_preparation_TQ}, Matrix Product States (MPS) \cite{MPS_periodic, MPS_preparation, MPS_Tensor_approx, quantum_state_preparation_TQ, TN_explanation_paper_1}, Riemannian Optimization \cite{riemannian, Riemannian_Manifolds}, tensor train decomposition (TT-decomposition) \cite{Tensor_decomposition_Reuben, Exponential_Machines}, for the implementation of the models.
Additionally, in Sec.~\ref{sec:Implementation}, we explain the architectures of the models, how we used the aforementioned techniques in the models, and how we implemented the VQC (in Sec.~\ref{sec:VQC_model}) and TN (in Sec.~\ref{sec:TN_model}) models in general. Furthermore, in Sec.~\ref{sec:Results} we present the experimental results and discuss the performance of both models under a common denominator. Finally, in Sec.~\ref{sec:Discussion}, we discuss the trade-offs between the models on binary classification tasks. Moreover, in the same section, we refer to potential future work and also strengthen our results by verifying the experiments.

\section{\label{sec:preliminaries} Preliminaries}

\subsection{\label{sec:ML} Quantum Machine learning}

QML can be characterized as the combination of classical ML with quantum mechanics to some extent. The quantum mechanical element can be inserted through different means. Some examples are the implementation of quantum data as qubits, the use of a parameterized quantum circuit (PQC) \cite{PQC}, for the training of the model, or even quantum algorithms such as Quantum Approximate Optimization Algorithm (QAOA) \cite{QAOA}, Variational Quantum Eigensolver (VQE) \cite{VQE} or Quantum Support Vector Machines (QSVM) \cite{QSVM}, for the further improvement of the model training and optimization \cite{QML_Challenges}.

A specific application of PQC is the variational quantum classifier (VQC) \cite{PQC, VQC}. The VQC is composed of two different parts, the encoder part, and the variational part \cite{PQC}. In the encoder, a quantum circuit is utilized to encode the features of samples into qubits. It is obvious that there are many different encoding techniques, such as angle encoding, amplitude encoding, wave-function encoding, and others \cite{data_encodings, PQC}. The choice of encoding is closely related to different kernel methods. These methods are used to project the data into a higher-dimensional feature space, where the same problem is typically easier to solve. So, the proper choice of feature space is paramount for the solution of the problem.

For example, non-linear feature maps \cite{non-linear_feature_maps, non-linear_features_QML} are capable to project data into a feature space where their relative distance is much different. In that way, it is possible that the samples can be easier distinguished from one another, and thus a binary classification among them can be achieved accurately. The inner product of two points (samples) in the feature space characterizes the kernel similarity function \cite{Practical_QML}.  

The second component of a VQC is the variational or parameterized part, which is a quantum circuit of a given ansatz. It generally consists of entangling layers and rotation gates on the qubits, with free, tunable parameters. The main hyper-parameter of this model is the number of variational layers chosen for the variational architecture. In order to give some depth to the variational circuit, which is crucial for its performance, we have to repeat the structure of $CNOT$ gates and rotations multiple times. Finally, measurements on one or more qubits are made, so the model can make predictions after observing the output states.

\subsection{\label{sec:TN} Tensor networks}

A tensor network can be described as the diagrammatic representation of a collection of tensors \cite{Patrick_tensors, bubbling}. Tensor networks are usually used in many-body quantum systems \cite{intro_to_TN_springer, tensor-networks_geometry}, and one of the most well-studied families of tensor networks are matrix product states or MPS \cite{Florian_tensors, MPS_preparation}. 

\begin{figure}[h!]
    \includegraphics[scale=.5]{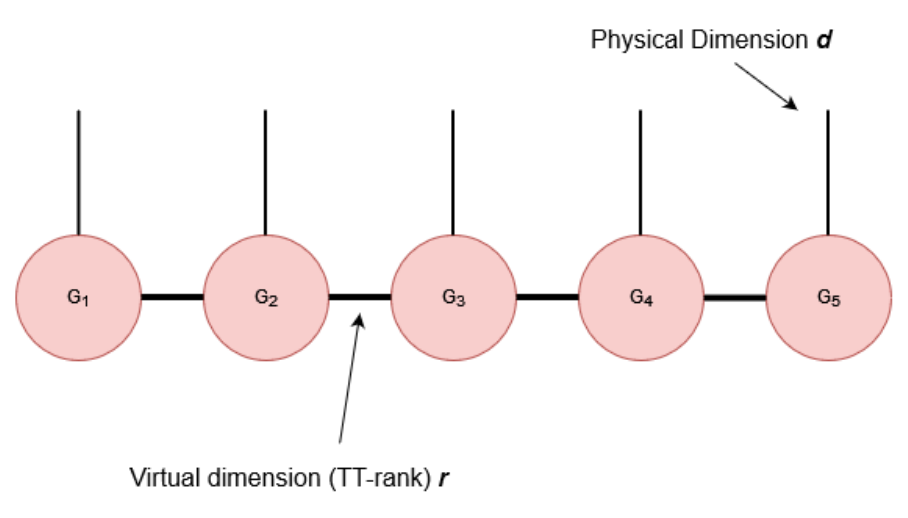}
    \caption{MPS example with five tensors $G_i, i \in \{1,2, \ldots ,5\}$, a common TT-rank $r_1 = r_2 = \ldots = r_4 = r$ and a common physical dimension $d_1 = d_2 = \ldots = d_5 = d$.}
    \label{fig:mps_structure}
\end{figure}

An MPS or a tensor-train (TT) is a $1$-dimensional state of a tensor network, where tensors are connected through one index called bond index (the dimension of which is called a virtual dimension or \textbf{tensor train-rank} \cite{Patrick_tensors}) and have another index called visible index, sticking out of the tensor (the dimension of which is called the physical dimension \cite{Patrick_tensors}) (Fig. \ref{fig:mps_structure}). In general, the TT-rank $r$ can vary from bond index to bond index and is regarded as a hyper-parameter of the model.

\begin{figure*}[ht!]
    \centering
    \includegraphics[width=1\linewidth]{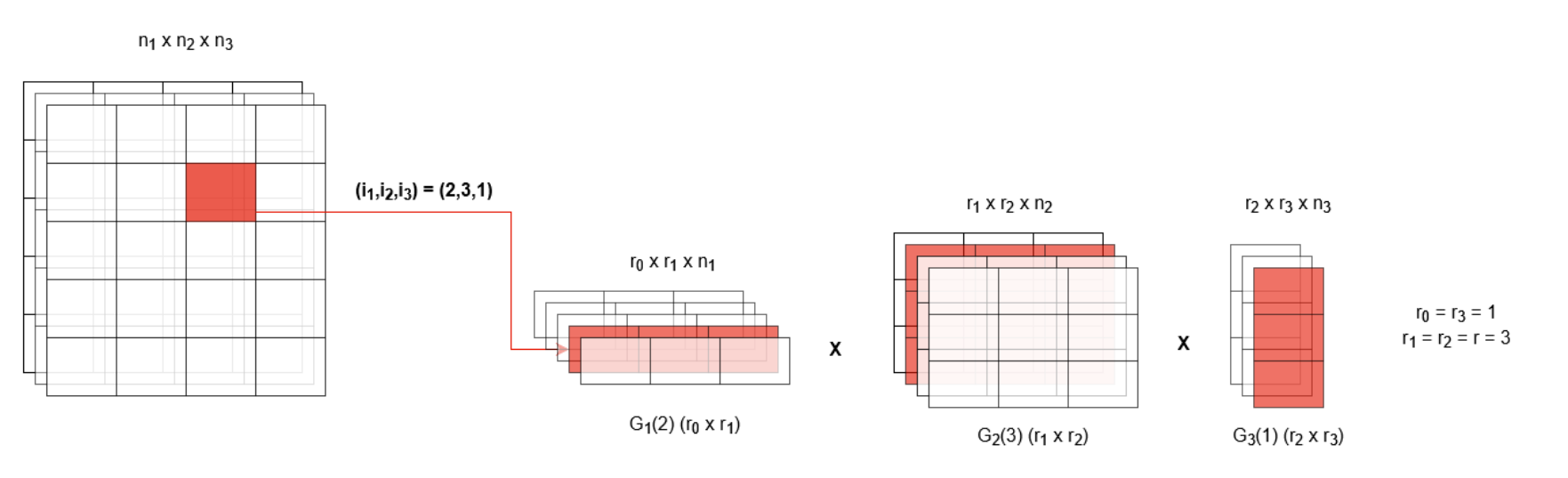}
    \caption{TT-decomposition example of a three-dimensional tensor of which the entry $\mathcal{A}_{2,3,1}$ is calculated. A fixed rank $r=3$ is used for all TT-cores $G_k(i_k)$, where $k \in \{1,2,3\}$ and $i_k$ is the respective index of the desired entry. So, in order to calculate the entry $\mathcal{A}_{2,3,1}$, we need to multiply the second matrix from the first TT-core (this corresponds to the first coordinate of the entry), with the third matrix of the second TT-core (which corresponds to the second coordinate of the entry) and finally multiply with the first matrix of the final TT-core (which corresponds to the third coordinate of the entry).}
    \label{fig:TT-decomposition}
\end{figure*}

After applying the TT-decomposition \cite{Tensor_decomposition_Reuben, Exponential_Machines} method on a tensor $\mathcal{A}$ (Fig.~\ref{fig:TT-decomposition}), the resulting network has the structure of an open boundary condition MPS. TT-decomposition is a way of manipulating multi-dimensional tensors without suffering from the curse of dimensionality \cite{curse_of_dimensionality, Tensor_decomposition_Reuben}. In the heart of TT-decomposition lies the singular value decomposition (SVD) \cite{SVD, HOSVD, Tensor_decomposition_Reuben}, which is applied $d-1$ times on the desired $d$-dimensional tensor $\mathcal{A}$, in order to get the $d$ tensors (TT-cores), which can efficiently represent $\mathcal{A}$. 

The general idea of TT-decomposition is to approximate all entries of the tensor $\mathcal{A}$ by the product of $d$ matrices $G_k(i_k)$ which belong on $G_k$ TT-cores of dimension $r_k \times r_{k-1}$ (with $r_0 = r_d = 1$ since we want the product to return a scalar value), within an error $\epsilon$. The indices $i_k$, with $k \in \{1,2 \ldots ,d\}$ and $i_k \in \{1,2, \ldots, n_k\}$, represent the dimension which enumerates over the $k$-th index of tensor $\mathcal{A}$, where $n_k$ is the dimension of the $k$-th index of $\mathcal{A}$. More specifically, we have

\begin{equation}
    A_{i_1i_2\ldots i_d} = \prod_{k=1}^d G_k(i_k)
\end{equation}

In order to train a tensor $\mathcal{B}$ to approximate the tensor $\mathcal{A}$ through TT-decomposition, we need a proper optimizer. A suitable optimizer for this task is the Riemannian optimization algorithm \cite{ Exponential_Machines,Tensor_decomposition_Reuben}.

Riemannian geometry is a branch of differential geometry that includes and describes the Riemannian manifolds. Manifolds are topological spaces that locally resemble Euclidean spaces \cite{Riemannian_Manifolds}. A Riemannian manifold $\mathcal{M}$ is a smooth Hausdorff and second countable manifold (by smooth we mean it is $C^\infty$, infinitely times differentiable), equipped with a positive-definite smoothly varying inner product $g$ metric, which can be used to determine an inner product on each point $p$ of the tangent space $T_p\mathcal{M}$ of $\mathcal{M}$ \cite{riemannian,Exponential_Machines}.

In a given $d$-dimensional tensor $\mathcal{A}$, we can apply the TT-decomposition with a fixed rank $r_i = r, \: i \in \{1,2,\ldots,d-1 \}$ and of course $r_0 = r_d = 1$. All such tensors like $\mathcal{A}$, belong to a Riemannian manifold  

\begin{equation}
    \mathcal{M}_r = \{\mathcal{A} \in \mathbb{R}^{n_1 \times n_2 \times \ldots \times n_d}: \text{TT-rank}(\mathcal{A}) = r\}
\end{equation}

\begin{figure*}[ht!]
    \centering
    \includegraphics[width=1\linewidth]{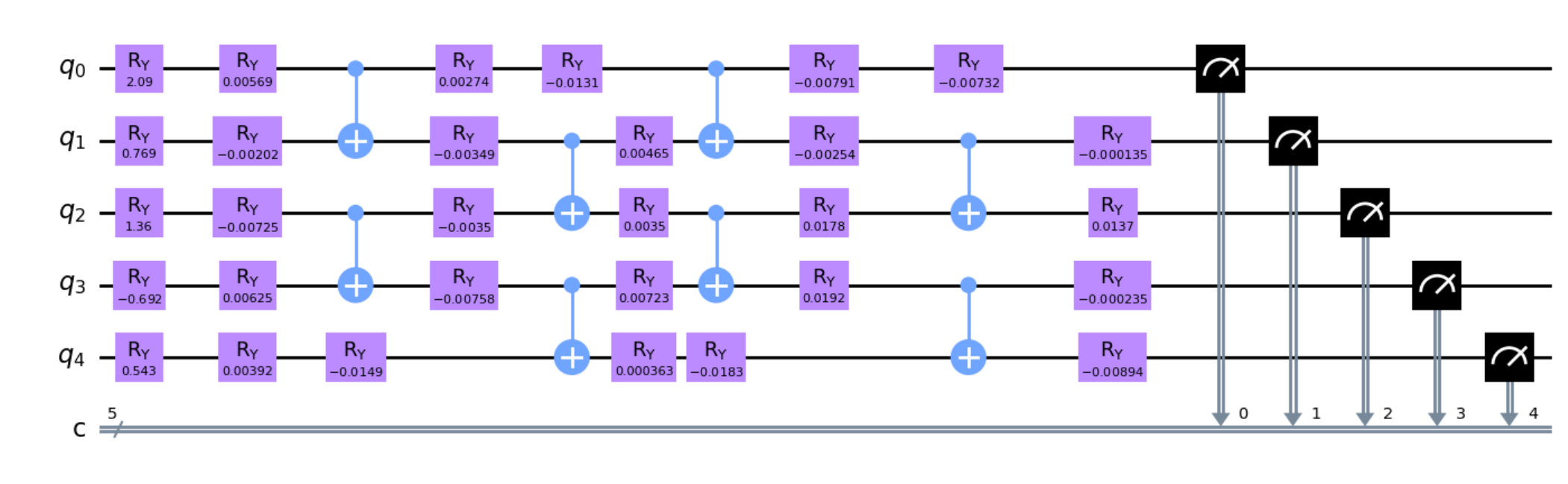}
    \caption{Example of the architecture used for the VQC model, with $5$ qubits and $2$ variational layers.}
    \label{fig:VQC_example}
\end{figure*}

Consider tensors $\mathcal{X}$ and $\mathcal{W}$, where $\mathcal{X}$ is of rank $1$ and consists of all the features that describe the data. $\mathcal{W}$ can be considered as a $d$-dimensional tensor, which contains all tunable weights of the model. As we show, the dimensions of $\mathcal{W}$ can be $n_1 = n_2 = \ldots = n_d = 2$ and thus its entries are in $\mathbb{R}^{2\times 2 \times \ldots \times 2}$. That is because the interactions of the features of the data can be multiplied with the weights according to the following equation:

\begin{equation}\label{eq:TN_predictor}
    \hat{y}(\mathbf{x}) = \sum_{i_1 = 0}^1 \sum_{i_2=0}^1 \ldots \sum_{i_d=0}^1 \mathcal{W}_{i_1i_2 \ldots i_d} \prod_{j=1}^d x_j^{i_j}
\end{equation}

Here $\hat{y}(\mathbf{x})$ represents the prediction of the model \cite{Exponential_Machines}. In other words, that prediction is the product of the relative weight with the features. When an entry of the weight tensor is multiplied with some features, then the relative indices of the weight tensor become $1$, and the remaining become $0$. 

For a two-dimensional example 

\begin{equation}
    \hat{y}(\mathbf{x}) = \mathcal{W}_{00} + \mathcal{W}_{10}x_1 + \mathcal{W}_{01}x_2 + \mathcal{W}_{11}x_1 x_2
\end{equation}

we see that for indices $i_1 = i_2 = 0$ we get $x_1^0 = x_2^0 = 1 $, for $i_1 = 1, i_2 = 0$, we get $x_1^1 = x_1, x_2^0 = 1$, and so on.

Riemannian optimization attempts to minimize the following loss function

\begin{equation}
    L(\mathcal{W}) = \sum_{s=1}^N MSE(\hat{y}(\mathbf{x}^{(s)}),y^{(s)}) + \frac{\lambda}{2} ||\mathcal{W}||_F
\end{equation}

where the term $\frac{\lambda}{2}||\mathcal{W}||_F$ is the $L_2$ regularization term, with $\lambda$ being the regularization parameter and $||\mathcal{W}||_F$ the Frobenius norm of the weight tensor \cite{frobenious_norm}. The $MSE(\hat{y}(\mathbf{x}^{(s)}),y^{(s)}):\mathbb{R}^2 \rightarrow \mathbb{R}$ is the mean squared error or the squared loss of the predictions $\hat{y}(\mathbf{x}^{(s)})$ towards true values $y^{(s)}$, and the $s$ refers to the current sample examined, with the total number of samples in the data being $N$.

Applying the Riemannian optimization algorithm on $\mathcal{W}$ we can fine-tune its entries in such a way that the model can make accurate predictions. The steps of the Riemannian optimization algorithm we follow are the calculation of the gradient of the loss, with respect to $\mathcal{W}$. Then, project it to the tangent space of $\mathcal{M}_r$. The projection at point $\mathcal{W}$ which is $T_\mathcal{W}\mathcal{M}_r$. We define the projection as 

\begin{equation}
    \mathcal{P} = P_{T_\mathcal{W}\mathcal{M}_r}\left( \frac{\partial L}{\partial \mathcal{W}}\right)
\end{equation}

Following the direction of the projection $\mathcal{P}$ with a small learning step $\alpha$ we get out of $\mathcal{M}_r$ and onto the tangent space $T_\mathcal{W}\mathcal{M}_r$. As a consequence, there is an increase in the TT-rank. To return to $\mathcal{M}_r$, we need to reduce (rounding \cite{Tensor_decomposition_Reuben}) the TT-rank back to $r$. This reduction can be achieved by retracting the projection from point $\mathcal{W}$, by a small learning step $\alpha$. In total, we take a step $\mathcal{W}-\alpha \mathcal{P}$ in order to return back to the manifold.

By recursively applying those steps, one can minimize the loss, and as a result, tune $\mathcal{W}$ to minimize the $L(\mathcal{W})$.

\section{\label{sec:Implementation} Implementation}

 We compare the performance on a binary classification task between a Variational Quantum Classifier (VQC) and a Tensor Network (TN) with Riemannian optimization. We applied both models on the UCI car dataset of 2013 \cite{car_data}. 
 
 The dataset has $1728$ samples with six categorical features each. Converting the categorical features to binary with one-hot encoding we end up with $21$ binary features in total. For the experiments, we used a random splitting of the data, with a splitting ratio of $80\%$ between training and validation sets. Additionally, for the data pre-processing we used PCA, so we were able to run experiments with $2,5,10$ and $16$ principal components of the data, as well as all $21$ features. 

The dataset originally was a multi-class classification problem, with classes (unacc, acc, good, vgood), which refers to the acceptability of each car in the dataset. In order to convert it to a binary classification problem, we merged all three classes (acc, good, vgood) into one class (namely "acc") which consists $29\%$ of the data, we later represented this class with $+1$, and the other class ("unacc") with $-1$ which consists the $71\%$ of the data. The dataset can be characterized as unbalanced and thus the choice of accuracy as a metric might be incorrect. However, after extended experimentation, both accuracy and F1-score, which is an excellent metric for imbalanced datasets, reported almost the same results \cite{Classification_metrics, Metrics_and_MSE}. The features of the dataset consist of the buying price, price of maintenance, number of doors, number of people to carry, size of luggage boot, and estimated safety of the car. 

We initialize both models with random weights and used the validation accuracy as a comparison metric between them.  

\subsection{\label{sec:VQC_model} Variational Quantum Classifier model}

For the implementation of the encoding part of VQC, we used the cosine/sine encoding \cite{Cos_sin_encoding,cos_sin_encoding_2}. For this encoding, we start from an all $\ket{0}$ state and we apply $R_y$ single qubit rotations on every qubit. After the encoding part, each qubit would be in the state

\begin{equation}
    \ket{0} \rightarrow R_y\ket{0} = \cos{\left(\frac{\pi}{2}x_k\right)}\ket{0} + \sin{\left(\frac{\pi}{2}x_k\right)}\ket{1}
\end{equation}

where $k \in \{1,2,\ldots, N\}$ and $N$ is the total number of features/qubits used for each sample. In that way, we introduce and pass information from the data to the quantum circuit. 

After the encoding part, we have to establish a variational architecture. We chose to follow the Noisy Intermediate Scale Quantum (NISQ) friendly architecture used in \cite{quantum_state_preparation_TQ}. The main hyper-parameter here is the number of variational layers used. Each layer consists of $CNOT$ gates with control on all odd-numbered qubits, then a layer of $R_y$ rotations applied on each individual qubit, $CNOT$ gates with control on every even-numbered qubit, and one more $R_y$ layer on every qubit. In total, every layer needs to train $2N$ parameters (plus $N$ for the $0$-th layer, which consists of only an $R_y$ rotation layer on every qubit). Thus, $N(2L+1)$ parameters need to be tuned, where $L$ is the number of layers used.

In Fig.~\ref{fig:VQC_example} an example of the architecture of the VQC model is illustrated. This example refers to a five-qubit model with two variational layers. At the end of the circuit, we utilize only the first measurement for the prediction. If the expectation value of the measurement falls under the predefined threshold, we classify the sample to the $-1$ ("unacc") class, otherwise, we classify it to the $+1$ ("acc") class. The threshold we use throughout all the experiments is set to $0.5$.

The MSE has been used as a loss function for VQC and for the circuit implementation Pennylane (version = 0.30.0) library has been used \cite{Pennylane}. For the optimizer, we tried Standard Gradient Descent (SGD), Nesterov momentum optimizer \cite{momentum_opt}, and Adaptive Momentum optimizer (Adam) \cite{adam, adagrad}. Adam returned the best results, so we chose that for all the experiments with VQC. During training, a batch size of $32$ samples was used, along with a decaying learning rate with an initial value of $0.1$ and a decay rate of $0.95$ \cite{decaying_learning_rate}.

\subsection{\label{sec:TN_model} Tensor Network model}

For the Tensor Network model (TN), we followed the implementation used in \cite{Exponential_Machines}. In this model, the method of TT-decomposition has been applied in order to manipulate the weights tensor in a more efficient way, especially when working with high-dimensional data. Polynomial encoding as defined in \eqref{eq:TN_encoding} has been used as data encoding.

\begin{equation}\label{eq:TN_encoding}
    \mathcal{X}_{i_1i_2\ldots i_d} = \prod_{k=1}^d x_k^{i_k}
\end{equation}

where, $\mathcal{X}_{i_1i_2\ldots i_d}$ represents the tensor which includes the features of the samples, and $x_k^{i_k}$ is the $k$-th feature of the sample. For the values $i_j$, the indices $j \in \{1,2,\ldots, d\}$ and $i_j \in \{0,1\}$. Thus, for the feature of a sample, it will be

\[   
x_j^{i_j} = 
     \begin{cases}
       1 &,\text{if} \: i_j = 0\\
       x_j &, \text{if} \: i_j = 1 
     \end{cases}
\]

This model is trained with Riemannian optimization and uses the logistic loss as a loss function. Its main hyper-parameter is the TT-rank used for the TT-decomposition. The total number of parameters that need to be trained in this model is $2Nr^2$, where $N$ is the number of features used, and $r$ is the TT-rank. We notice that the number of parameters in the TN model scales much faster in comparison with VQC since the TN parameters scale quadratically with the rank. 

For the implementation of the TN model, the ttpy library, which is a python implementation of the TT-Toolbox library \cite{tt-toolbox}, has been used, in coordination with other libraries with mathematical tools, such as numpy \cite{numpy} and scikit-learn \cite{scikit-learn}.

As mentioned above, the predictor of this model is based on the simple linear product between the features tensor $\mathcal{X}$ and the weights tensor $\mathcal{W}$. So in total, we can rephrase equation \eqref{eq:TN_predictor} as

\begin{equation}
    \hat{y}(\textbf{x}) = \langle \mathcal{X},\mathcal{W} \rangle
\end{equation}

There is no need for a separate bias term since this is integrated in $\mathcal{W}$ as $\mathcal{W}_{00\ldots 0}$.

\section{\label{sec:Results} Results}

\begin{figure}[h!]
    \centering
    \includegraphics[width=\columnwidth]{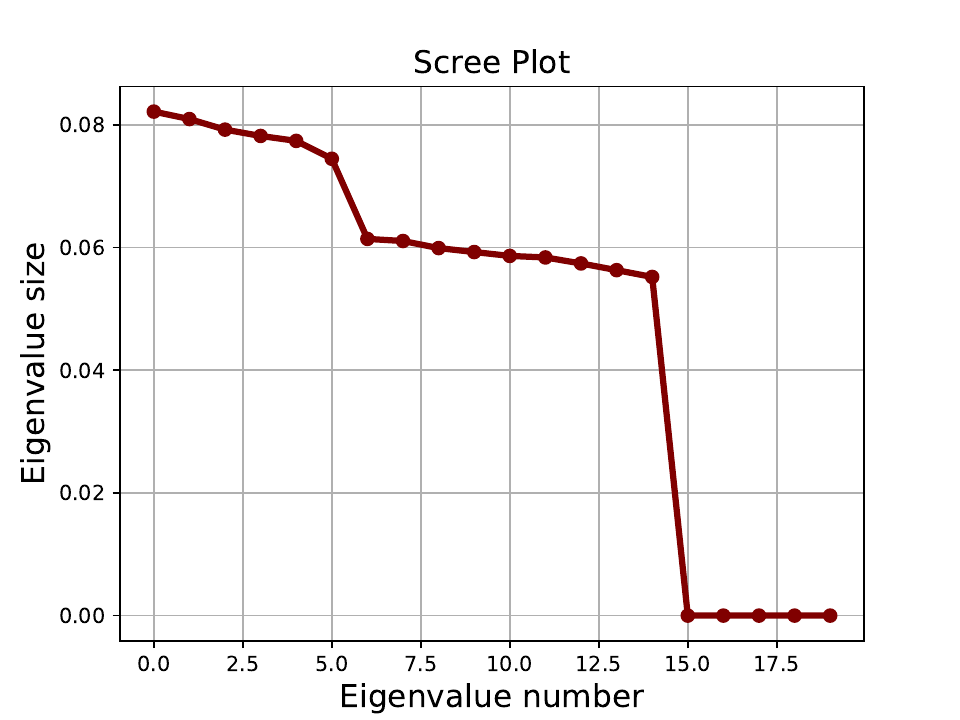}
    \caption{Scree plot analysis for up to $20$ principal components. The eigenvalue number represents the principal components starting from $0.0$ which corresponds to the $1$st principal component, and the eigenvalue size represents the percentage of the original data that each principal component expresses.}
    \label{fig:scree}
\end{figure}

We compare the VQC and TN models with different numbers of qubits and principal components. In order to decide which number of principal components we will run experiments for, we compose a scree plot analysis \cite{scree} in Fig.~\ref{fig:scree}.

\begin{figure}
\centering
  \includegraphics[clip,width=\columnwidth]{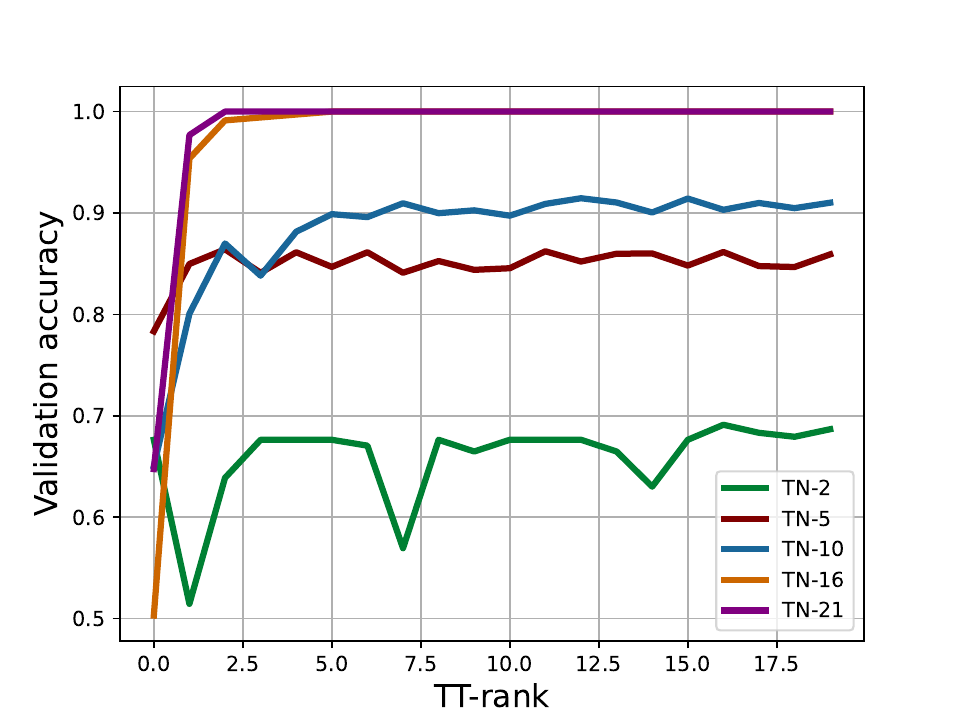}%
  \label{fig:TN_all}

  \includegraphics[clip,width=\columnwidth]{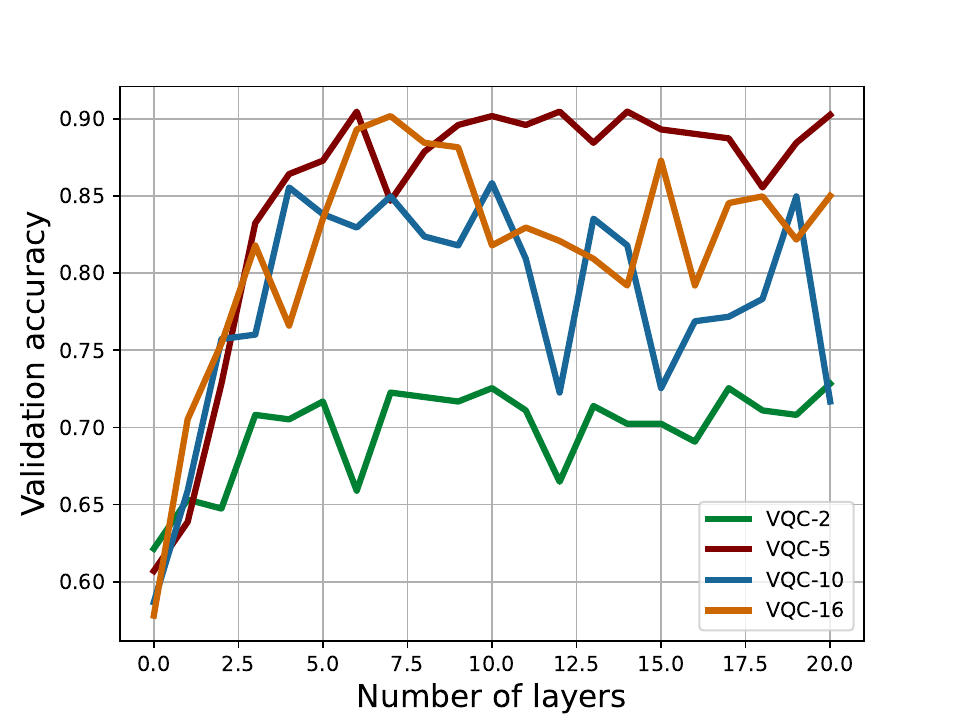}%
  \label{fig:QC_all}
\caption{ {\bf (a)} Plot of the validation accuracy for the Tensor networks model with the varying number of TT-rank.
{\bf (b)} Plot of the validation accuracy for the Quantum circuit model with the varying number of variational layers.
Colours inidcate different numbers of principal components/qubits ($2$, $5$, $10$, $16$).}
\label{fig:vqctn}
\end{figure}

Each point in Fig.~\ref{fig:scree} represents the percentage that is going to be added to the total expressibility of the dataset. As a result, it is logical to make experiments with $2$ qubits, which seem to give low expressibility, $5$ qubits which are again in low expressibility but still have to add to the overall eigenvalue size, $10$ qubits which are in the mid-range of the eigenvalue size contribution, and $16$ qubits which in principle add almost $0\%$ to a model with $15$ principle components. 

In Fig.~\ref{fig:vqctn}, we show the results of a binary classification experiment on the UCI car dataset using a VQC in comparison to a TN model as described above. In both experiments, we used different numbers of principal components from the car dataset and recorded their accuracy with the number of variational layers/TT-ranks and the number of principal components used in training. As a result, it seems that for the tested dataset and model architectures, the TN model increases its performance significantly as the number of principal components increases. Additionally, VQC also increases its performance and achieves better classification, however with more principal components, the training time, and thus the effort for parameter optimization in the variational layer increases significantly and often leads to local minima or vanishing gradients.

\begin{figure*}
\centering
\includegraphics[scale=0.5]{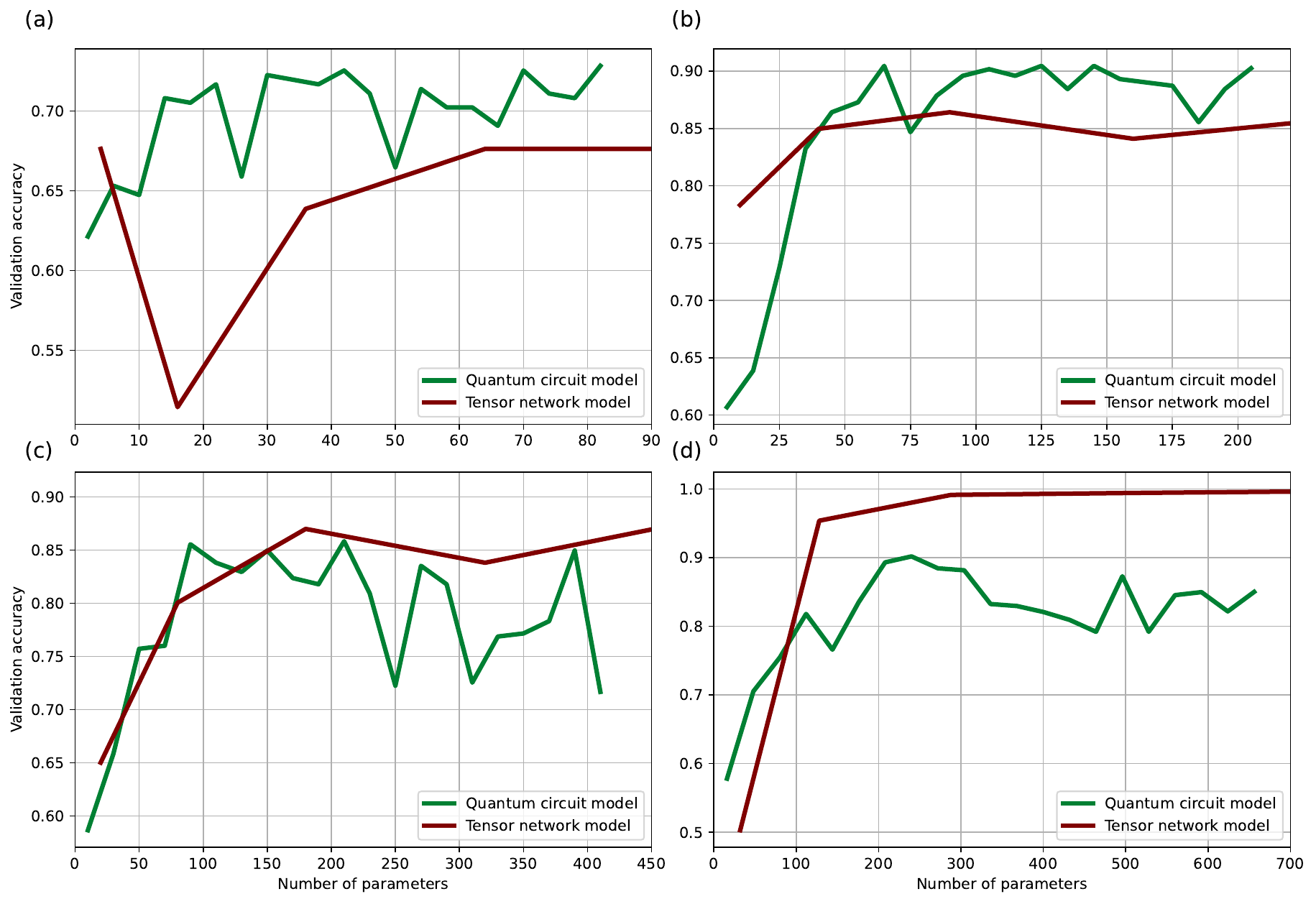}
\caption{The validation accuracy achieved by TN and VQC models depending on the number of trainable parameters. {\bf (a)} Comparison of validation accuracy of for the $2$ qubits models where VQC uses significantly less parameters. {\bf (b)} Comparison when TN and VQC utilize $5$ qubits, still VQC dominates in overall accuracy. {\bf (c)} Comparison with $10$ qubits models, TN starts to outperform VQC. {\bf (d)} In comparison with $16$ qubits, TN clearly outperforms VQC when uses over $100$ trainable parameters.}
\label{fig:tn_vqc_param_comparison}
\end{figure*}

In Fig.~\ref{fig:tn_vqc_param_comparison}, we show the respective plot of the validation accuracy with the number of trainable parameters, for $5$ qubits. It is evident that VQC performs better than TN. As the trainable parameters increase, notice that the performance of TN converges, in contrast with VQC which oscillates around $90\%$ but always above TN.

In Fig.~\ref{fig:tn_vqc_param_comparison}, we compare the $10$ qubits TN and VQC. Notice that TN surpasses VQC considering it achieves better accuracy than VQC for $\geq 150$ trainable parameters. TN performance in $10$ qubits seems comparable with the $5$ qubits VQC. However, the experiments with more than $10$ qubits, indicate that TN is the dominant model for high-dimensional binary classification tasks. This observation is based on the poor trainability of VQC for a large number of qubits.

For $16$ qubits in Fig.~\ref{fig:tn_vqc_param_comparison}, TN is clearly better than VQC, regarding the validation accuracy. TN manages to achieve perfect classification and VQC shows a more stable behavior in comparison with the $10$ qubit experiment. Under $100$ parameters, VQC surpasses TN, which is not the case for more than $100$ parameters.

\section{\label{sec:Discussion} Discussion}

 We are not able to claim that one model is better than the other. Accuracy depends on the available computational power and time. Also, the dimensionality of the data plays a significant role when we want to choose between the models. If the only consideration is the overall accuracy, then TN models might be a better choice, especially, if the problem involves high-dimensional data, to give a perspective, data with more than $10-15$ features. TT-decomposition enables us to manipulate the high-dimensional weight tensor much faster than other methods. Riemannian optimization seems to be an excellent fit as an optimizer to TN since it outperforms SGD \cite{Exponential_Machines} and assists TN in achieving better classification. A drawback of TN is that requires significantly more trainable parameters compared to VQC.

On the other hand, VQC equipped with a dimensionality reduction technique (such as PCA), or used on a problem with low-dimensional data, might be a better choice than TN. For example, $5$ qubits VQC can reach $\sim 91\%$ validation accuracy on the UCI car dataset. We observed that VQC achieved maximum accuracy after a small number of epochs. This is especially promising to reduce the training time of VQCs with a larger number of qubits. 

There are indications that VQC can achieve even better accuracy with more qubits. As shown in Fig.~\ref{fig:QC_all} with $16$ qubits, it is capable of outperforming the $5$ qubits for $6-8$  layers. So with hyper-parameter tuning or more training, VQC might also be suitable for high-dimensional data. However, because training time for the $16$ qubits VQC requires much computation time, we did not investigate this possibility as much as it seems to deserve.

Overall, by examining the plots, we notice that VQC is consistently the model that achieves the best validation accuracy for a small number of variational layers or low TT-rank. This result strengthens our previous finding about VQC, being the preferable method when we need to train a model within a limited time frame.

\newpage

\nocite{*}

\bibliography{lib}

\end{document}